\documentclass[preprint,aps,amssymb]{revtex4-1}
\usepackage{graphicx}
\usepackage{dcolumn}
\usepackage{color}

\newcommand{\beginsupplement}{%
        \setcounter{table}{0}
        \renewcommand{\thetable}{S\arabic{table}}%
        \setcounter{figure}{0}
        \renewcommand{\thefigure}{S\arabic{figure}}%
     }

\usepackage[fleqn]{amsmath}

\usepackage{epstopdf}

\def\Vg{V_G}
\def\Ek{E_0}
\def\dE{\delta E}
\def\Ic{I_{C}}
\def\vf{v_F}
\def\kb{k_B}
\def\Ib{I}
\def\Vb{V_B}

\begin{document}

\title {Ballistic graphene Josephson junctions from the short to the long regime}

\author {I. V. Borzenets$^{1}$$^{*}$, F. Amet$^{2}$, C. T. Ke$^{3}$, A. W. Draelos$^{3}$, M. T. Wei$^{3}$, A. Seredinski$^{3}$, K. Watanabe$^{4}$, T. Taniguchi$^{4}$, Y. Bomze$^{3}$, M. Yamamoto$^{1,5}$, S. Tarucha$^{1,6}$, G. Finkelstein$^{3}$}

\affiliation{
$^{1}$Department of Applied Physics, University of Tokyo, Bunkyo-ku, Tokyo 113-8656, Japan.
\\$^{2}$Department of Physics and Astronomy, Appalachian State University, Boone, NC 28607, USA.
\\$^{3}$Department of Physics, Duke University, Durham, NC 27708, USA.
\\$^{4}$Advanced Materials Laboratory, National Institute for Materials Science, Tsukuba, 305-0044, Japan.
\\$^{5}$PRESTO, JST, Kawaguchi-shi, Saitama 332-0012, Japan.
\\$^{6}$Center for Emergent Matter Science (CEMS), RIKEN, Wako-shi, Saitama 351-0198, Japan.
\\$^{*}$Correspondence should be sent to I.V.B. (email: ivan@meso.t.u-tokyo.ac.jp)
 }

\begin{abstract}
We investigate the critical current, $\Ic$, of ballistic Josephson junctions made of encapsulated graphene/boron-nitride heterostructures. We observe a crossover from the short to the long junction regimes as the length of the device increases. In long ballistic junctions, $\Ic$ is found to scale as $\propto \exp(-k_BT/\dE)$. The extracted energies $\dE$ are independent of the carrier density and proportional to the level spacing of the ballistic cavity, as determined from Fabry-Perot oscillations of the junction normal resistance. As $T\rightarrow 0$ the critical current of a long (or short) junction saturates at al level determined by the product of $\dE$ (or $\Delta$) and the number of the junction's transversal modes. 
\end{abstract}


\maketitle

Encapsulated graphene/boron-nitride heterostructures emerged in the past year as a medium of choice for studying proximity-induced superconductivity in the ultra-clean limit~\cite{Vandersypen2015,Geim2015,Yacoby,amet2016}. These junctions support the ballistic propagation of superconducting currents across micron-scale graphene channels, and their critical current is gate-tunable across several orders of magnitude. In these devices, a rich phenomenology arises from the interplay of superconductivity with ballistic transport~\cite{Vandersypen2015}, cyclotron motion~\cite{Geim2015}, and even the quantum Hall effect at high magnetic field~\cite{amet2016}. 

\begin{figure}[h!]
\includegraphics[width=0.5 \columnwidth]{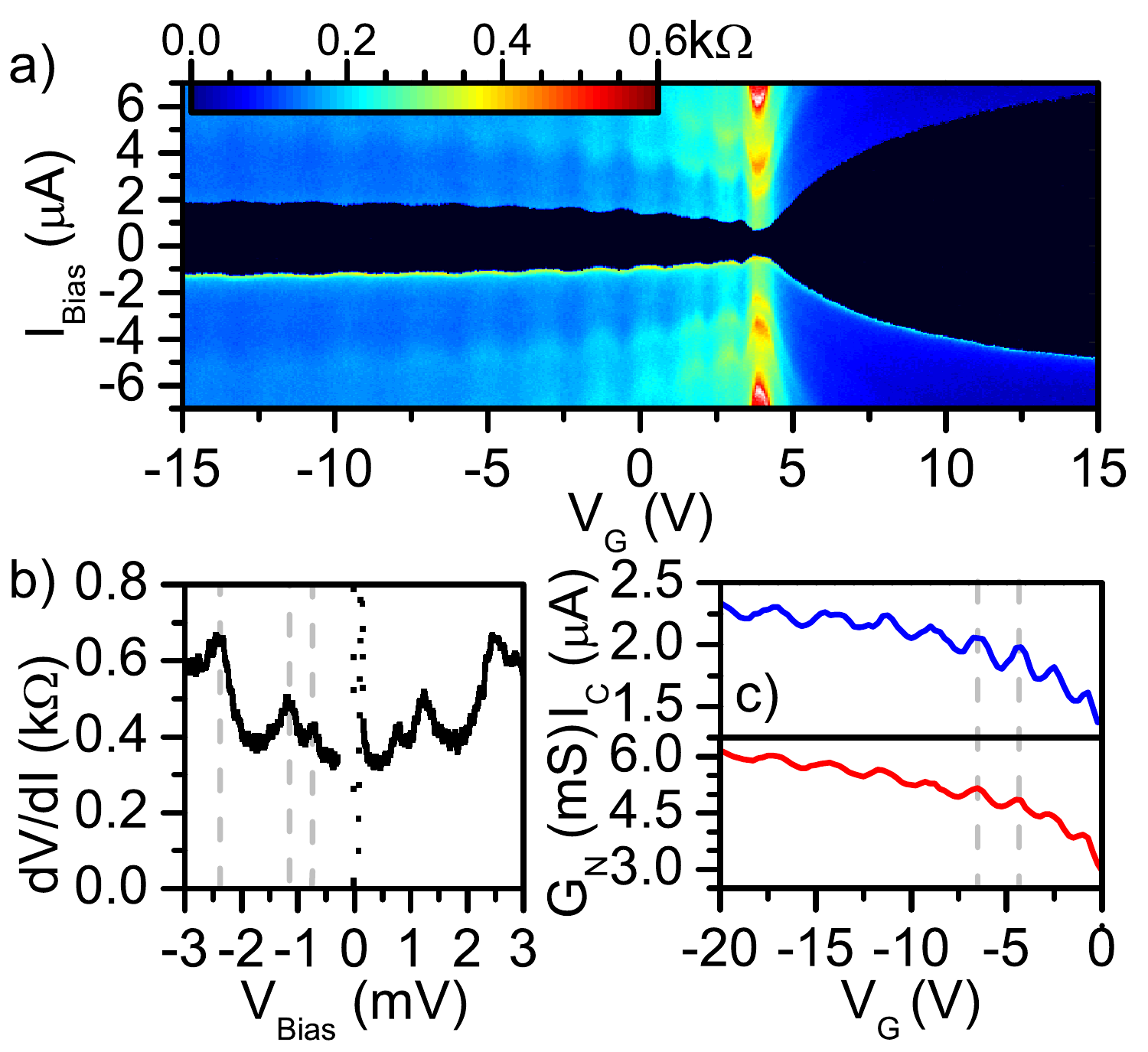}
\caption{\label{fig:overview}  a) Map of differential resistance versus current $\Ib$ and gate voltage $\Vg$. The data are shown for Junction A and taken at a temperature $T=1.5K$. The superconducting region of zero resistance can be observed around $\Ib=0$. The current through the junction is swept from negative to positive; therefore, the transition at the negative $\Ib$ corresponds to the retrapping current $I_R$, while the transition at the positive $\Ib$ corresponds to the switching current $I_S$. b) Differential resistance versus bias voltage ($V_B$) for Junction A taken at Dirac point. Several multiple Andreev reflection (MAR) peaks  are observed: $2\Delta$, $\Delta$, $2/3\Delta$; with $\Delta\approx1.2$meV. c) The critical current $I_C$ (top) and the normal conductance of the junction (bottom) plotted vs. gate voltage $V_G$ in the hole conduction regime. Both quantities demonstrate Fabry-Perot oscillations and are roughly proportional to each other. }
\end{figure}

In a superconductor - normal metal - superconductor (SNS) junction, single particles in the N region cannot enter the superconductor and therefore experience Andreev reflections at each S-N interface. This results in Andreev bound states (ABS), which are capable of carrying superconducting current across the N region. In long ballistic junctions, the energy spectrum of the ABS is quantized with a level spacing of $E_0 = \pi\hbar v_F/L$, where $L$ is the junction length and $v_F$ the Fermi velocity~\cite{Kulik,Bardeen,book,Tinkham,Golubov}. The energy of ABS cannot exceed the superconducting gap $\Delta$, so in the short junction regime, $L \lesssim \xi\equiv\hbar\vf/\Delta$, only a single ABS remains.

\begin{figure*}[ht!!]
\includegraphics[width=1 \columnwidth]{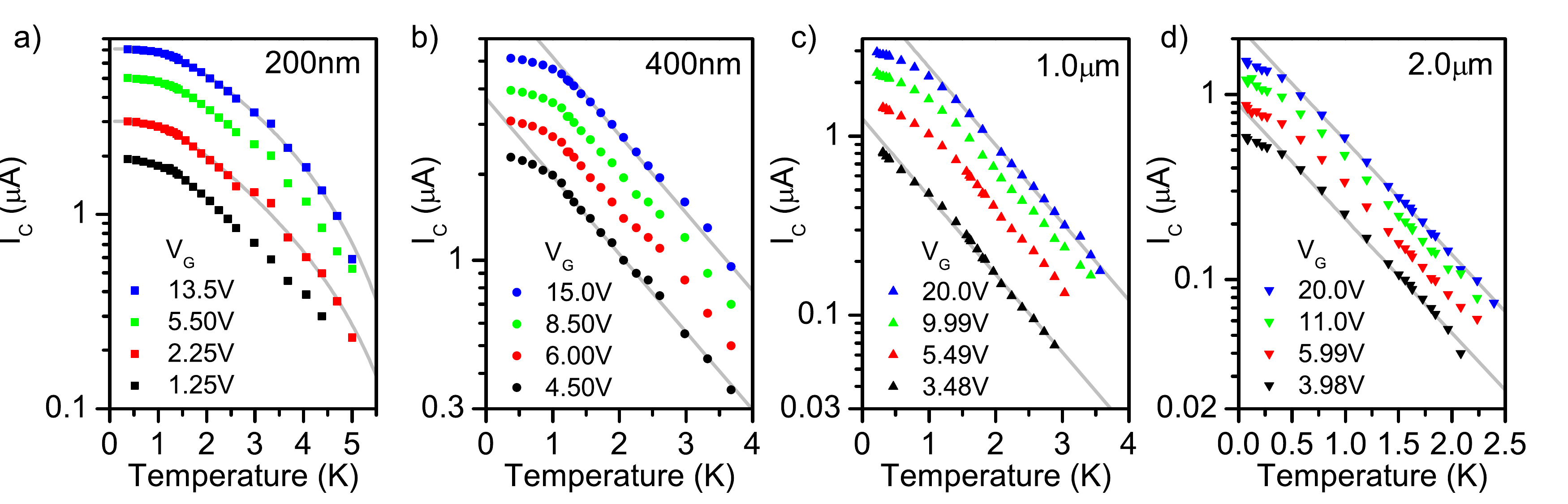}
\caption{\label{fig:overview} Critical currents $\Ic$ plotted on a semi-log scale versus temperature $T$ for Junctions A-D. Several gate voltages are presented for each junction;  the values of $V_G$ are shown relative to the Dirac point. a) The data for the shortest junction, A ($L=200$ nm). The gray lines are fitted according to eq. (1), using the superconducting gap $\Delta$ extracted in Figure 1b. b-d) $\Ic$ vs. $T$ for Junctions B-D respectively (see Supplementary for Junctions E, F, G). The slope of $\log{(\Ic)}$ vs. $T$ is independent of $\Vg$. In the case of long ballistic graphene junctions, the inverse slope $\dE$ is expected to be independent of the carrier density and inversely proportional to $L$. 
}
\end{figure*}

In this work we study several ballistic junctions of different length and demonstrate that the temperature dependence of the critical current dramatically differs in the long and short regimes. For long junctions, we observe an exponential scaling of the current through the junction $\Ic\propto \exp(- k_B T/\dE)$, where $\dE\approx\hbar v_F/2\pi L$~\cite{Kulik,Bardeen, Svidzinski1,Svidzinski2}. Note that in graphene $v_F$ is a constant, and $\dE$ is expected to be independent of the carrier density or the mobility (as long as the junction remains ballistic.) For comparison, in a short junction we observe a different scaling, as expressed in eq. (1), in excellent agreement with the theory~\cite{Beenakker92, Beenakker91, Lee}.

Our graphene layers are exfoliated from Kish graphite and encapsulated in hexagonal boron-nitride (hBN) using the ``pick-up'' method~\cite{Wang}. Heating beyond 250$^{\circ}$C causes bubbles of trapped adsorbates to migrate towards the edges of the graphene mesa, effectively cleaning it. The edges of the graphene flake are exposed by etching through the hBN-graphene-hBN stack with a CHF$_3$/O$_2$ plasma (flow rates $40/6$ sccm) at $1$Pa and $60$W power. The etching time varies depending on the thickness of the top hBN layer. We use DC magnetron sputtering to form Molybdenum-Rhenium alloy contacts ($50/50$ wt$\%$), with a measured superconducting gap $\Delta_0\approx 1.2$meV (Figure 1b).  These contacts are $100-120$nm thick and are deposited at a rate of $\sim50$nm/min (with a pressure of $2$mTorr and a power of $160$W~\cite{amet2016}). In this work we studied seven Josephson junctions with lengths ranging from 200$\,$nm to 2000$\,$nm. Device dimensions are listed in the supplementary information~\cite{suppl}. Junction A is found to be in the short regime, Junctions B and E are intermediate, while Junctions C, D, F, and G are in in the long regime. Below we present primarily the data measured on four junctions A-D ($L=$ 200nm, 400nm, 1$\mu$m, and 2$\mu$m) fabricated on the same substrate.

The junctions are measured in a four-terminal setup with the carrier density in graphene being controlled by a gate voltage, $\Vg$. Figure 1 presents a map of the differential resistance $dV/dI (\Vg, \Ib)$, measured on Junction A at $T=1.5$ K. The dark region of vanishing resistance indicates a supercurrent, which persists at all values of $\Vg$. As the current is swept from the negative to the positive values, the transition from the normal to the superconducting state is seen at negative bias when $\vert \Ib \vert =I_R$ (the retrapping current.) The transition from the superconducting back to the normal state happens at positive bias when $\Ib=I_S$. As commonly observed in graphene Josephson junctions, at low temperatures the samples exhibit hysteresis, $I_S \gtrsim I_R$~\cite{heersche_2007,du_2008,gueron_2009,23,24, Overheat}, which could be attributed to either underdamped junction dynamics~\cite{Tinkham,23}, or to the self-heating by the retrapping current~\cite{Pekola,Overheat}. As discussed in the supplementary material, the second scenario is more likely for most of the range studied here. Based on the measurements of the switching statistics~\cite{Ulas, HJ_Lee,Ting,suppl}, in the following we will use the switching current to represent the true critical current of the junction, $I_C$.




In the hole-doped regime, the reflections of ballistic charge carriers from the n-doped contact interfaces yield the quantum (``Fabry-Perot'') interference. A very similar oscillation pattern could be observed in the dependence of both the the normal conductance, $G_N$, and the critical current $I_C$ on gate voltage $V_G$ (Figure 1c)~\cite{Geim2015,Vandersypen2015,amet2016}.
Oscillations are also observed as a function of bias voltage $V_B$ (Figure 4a inset)~\cite{Young,Rickhaus,Geim2015,amet2016}.
 
The critical current $I_C$ is observed to rapidly decrease with temperature, however the functional form of $\Ic(T)$ strongly depends on the length of the junction. Figure 2 shows the evolution of $\Ic(T)$ from the short to the long regime. Each panel shows data measured for several values of $\Vg$, which from here on is shown relative to the Dirac point. The shortest junction (Figure 2a) can only support a single ABS; in this regime, the current  is:

\begin{equation}
\Ic(T,\phi) \propto \frac{e\Delta}{R_N} \frac{\sin{\phi}}{\sqrt{1-\tau\sin^2{\phi/2}}} \tanh{\bigg(\frac{\Delta}{2k_B T}\sqrt{1-\tau\sin^2{\phi/2}}\bigg)}
\end{equation}
where $\tau$ is the transmission coefficient of the S-N interface and $R_N$ the normal state resistance. For a given $T$, this expression should be maximized over $\phi$ to determine $\Ic(T)$~\cite{Beenakker91, Beenakker92, Lee}.  Moreover, at higher temperatures the superconducting gap will be suppressed;  we approximate the temperature dependence of the gap as $\Delta(T)\approx\Delta_0 \sqrt{1-(\frac{T}{T_C})^2}$, where $\Delta_0$ is the gap for $T\rightarrow0$, and $T_C$ is the critical temperature~\cite{Tinkham,Hagymasi,GapT}. Taking the complete temperature-dependent expression, we fit $I_C (T)$ for Junction A using the value $\Delta_0$$\,=\,$ 1.2 meV extracted from multiple Andreev reflections measurements.  The fit is in excellent agreement with the data (Figure 2a). 

\begin{figure}[h!!]
\includegraphics[width=0.5\columnwidth]{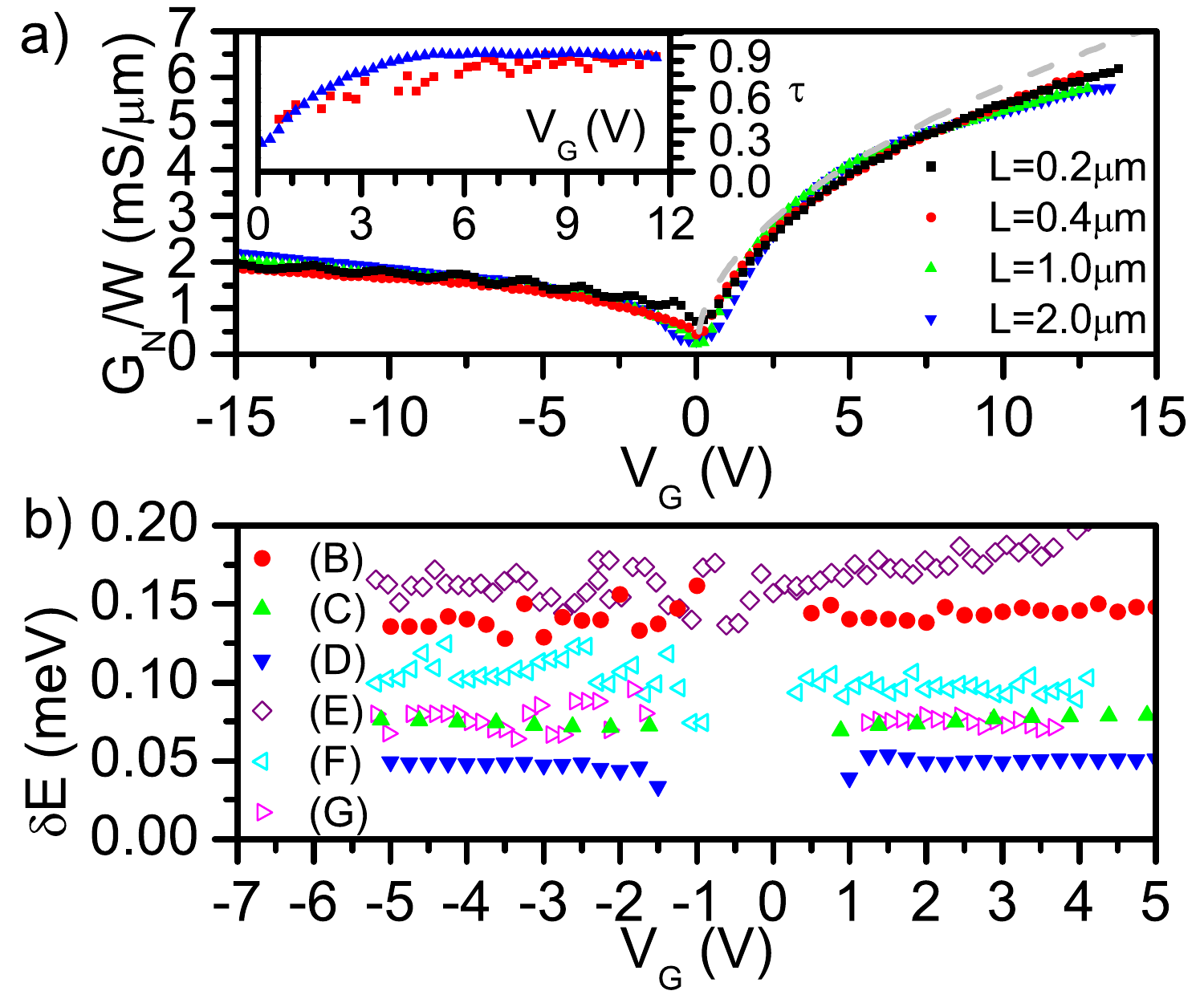}
\caption{\label{fig:overview}
a) Normal conductance of Junctions A-D normalized by the width of the junctions, $G_N/W$. Even though the device lengths are different by up to a factor of $10$, the three curves are very close to each other, thus proving the ballistic nature of these junctions. At positive $V_G$, $G_N$ of all junctions is found to approach $G_0=Ne^2 /h$ (gray dashed line), indicating consistently high contact  transparency for n-doping in these devices. Inset: Transmission coefficient $\tau$ of Junction A. $\tau$ is calculated via two methods: comparing the normal conductance $G_N$ to the ballistic limit $G_0=Ne^2/h$ (blue), and fitting the critical current $I_C$ vs. temperature $T$ (red). Both methods provide consistent results and indicate high contact transparency for N-doping. b) Energy $\dE$ extracted from the slope of $\log(\Ic)$ vs. $T$ for Junctions (B-G).  As expected in the long junction regime, $\dE$ depends only on device length $L$ and is almost density-independent through both the electron and hole doping. }
\end{figure}

The transmission coefficient $\tau$ extracted from the fit is plotted in the inset of Figure 3 as a function of the gate voltage. We can also estimate the transmission coefficient via an alternative method, by comparing the junction normal conductance $G_N$ to the ballistic limit of conductance, $G_0=Ne^{2}/h$, where $N=4\sqrt{\frac{n}{\pi}} W$ is the number of transversal modes and $n=V_G C_G/e$ is the carrier density. $\tau$ estimated as $G_N/G_0$ is shown in blue in the inset of Figure 3. Both methods provide consistent results, with $\tau$ in junction A reaching 90$\%$ at high densities. Furthermore, we find that the normal conductance of all junctions is very close to the ballistic limit. Figure 3a compares the normal conductance of junctions A-D normalized by junction's width (in fact, junctions A-C have the same widths). All four curves are very close to each other and approach the ballistic limit for positive gate voltages (dashed line). This result indicates two important facts: a) the contacts of all junctions are highly transparent on the n-doped side and b) the junctions' conductances do not depend on length, confirming their ballistic nature.

We now return to the critical current measured in the longer Junctions B-D. In Figure 2(b-d), $I_C$ is plotted on a semilogarithmic scale and clearly shows exponential dependence at high temperatures $T$ (over an order of magnitude in panels c and d). This is consistent with the expected long junction behavior $\Ic\propto \exp(- k_B T/\dE)$~\cite{Kulik,Bardeen,Svidzinski1, Svidzinski2, book,Golubov} and allows us to extract the energy scale $\dE$. 
The temperature dependence eventually saturates at low temperatures, when $\kb T$ becomes comparable to $\dE$. 

Figure 3b shows that for a given device $\dE(\Vg)$ remains roughly constant as a function of $\Vg$ for both electron and hole doping, as expected in the long ballistic regime. $\dE$ is on the order of 0.05 meV for the longest device, Junction D, and goes up to $\sim0.2$ meV for Junction E. While $\dE$ is consistent with the expected value of $\frac{\hbar v_F}{2\pi  L}$ for Junction D, it is suppressed for shorter junctions. As the devices are ballistic, the suppression of $\dE$ cannot be explained by the effective lengthening of the carrier path due to diffusion. 

To explain the suppressed $\dE$, we observe that the previous discussion of the long junctions neglected the coherence length $\xi$ compared to $L$. Taking $\xi$ into account suppresses the level spacing, which becomes $E_0 = \frac{\pi\hbar v_F}{L+\xi}$~\cite{Bagwell}. 
While the general expression for $I_C(T)$ in the $L \approx \xi$ regime is not known, numerical simulations show that it still roughly follows the $\propto \exp(- k_B T/\dE)$ dependence, with $\dE$ suppressed by a factor of $\sim 2$ compared to the estimate that neglects $\xi$ (Figure 3b in Ref.~\cite{Hagymasi}). In our case, $\xi \approx 550$ nm, which explains the suppressed $\dE$ in the intermediate regime (Junctions B, E). Eventually, the junction transitions to the short regime, where the exponential dependence no longer holds.

\begin{figure}[h]
\includegraphics[width=0.5 \columnwidth]{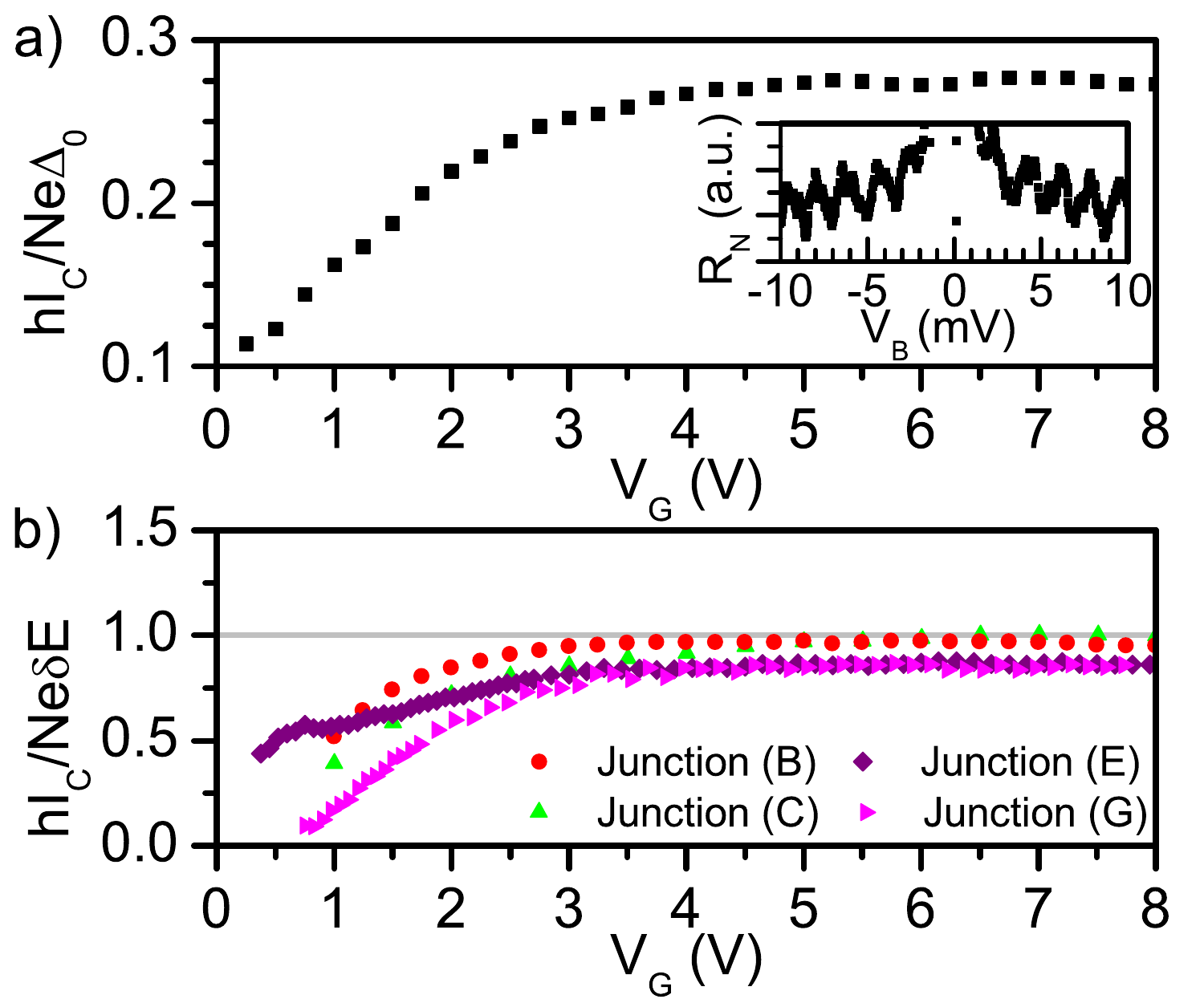}
\caption{ a) The ratio $h\Ic/N e \Delta_0$ measured on Junction A at $T=300$mK as a function of $V_G$. The number of modes $N$ is $W(4\sqrt{n/\pi})$, where $n$ is the carrier density as determined from $V_G$. As the gate voltage increases and the transmission of the graphene-MoRe interfaces approaches 1, the plotted ratio saturates.  Inset: Differential conductance versus bias voltage $\Vb$ for the 650 nm long Junction F, gated to the p-doped regime ($\Vg=-4.2V$). The period of the Fabry-Perot oscillations yields the level spacing, $\Ek \approx 2$ meV, which is consistent with the expected $\Ek=2\pi^2\dE$ ($\dE \approx 0.1$ meV for this junction.)   b) The ratio $h\Ic/N e \dE$ for junctions in the long regime (B,C,E,G), measured as a function of $\Vg$ at 60mK. (See supplementary for Junction F. Junction D does not yet saturate at the base temperature.) At higher $V_G$ the ratio  converges toward a constant value in all junctions. }
\end{figure}

We now turn to the saturation of $\Ic$ in the low temperature limit: $\kb T \ll \Delta_0$ for a short junction, or $\kb T \ll \dE$ for a long junction. 
In the long ballistic junction regime, the $T=0$ critical current is expected to be on the order of $e \dE/h$ per transversal mode~\cite{book,Golubov,Ishii}.  Figure 4b shows the ratio $\frac{h Ic}{N e \dE}$ as a function of the gate voltage. Strikingly, the curves for the four junctions are very close to each other and converge to a constant level of $ \approx 1$ at high gate voltage, where the graphene-MoRe interfaces are highly transparent. (See Supplementary for data on additional devices.) Similarly, the $T=0$ critical current per mode is expected to be $\sim e \Delta_0/h$ in an ideal short junction~\cite{Tinkham,Golubov}. Figure 4a plots the ratio $\frac{h I_C}{N e \Delta_0}$ for Junction A, which indeed saturates at high gate voltage, although its value $ \approx 0.3$ is significantly smaller than $\sim 2$ predicted by theory of Ref.~\cite{Titov}. Previous works have observed similar deviations from theory~\cite{Geim2015}. The mechanism for such suppression is unclear and can not be explained by environmental damping effects, nor the effect of imperfect transmission~\cite{suppl}. (Note: as there are currently no graphene-specific theoretical works predicting the ratio $\frac{h Ic}{N e \dE}$ in the long regime, it is unclear whether the value of $\sim 1$ observed in Figure 4b is coincidental.) 


The ratio $\frac{h Ic}{N e \dE}$ is significantly reduced close to charge neutrality. This suppression most likely arises from the $V_G$ dependence of the transmission coefficient $\tau$ of the superconductor-graphene interface. We extract the contact transparency from the junction normal resistance as $h/Ne^2R_N$ and find that while $\tau$ is close to 1 at high densities, it does get significantly suppressed close to the charge neutrality point. Considering this suppression allows us to partially account for the reduced $\frac{h Ic}{N e \dE}$ ratio (see Supplementary).

In conclusion, we studied the nature of the critical current in several ballistic superconductor-graphene-superconductor junctions. We find that in the short junction regime, $L \ll \xi$, the critical current follows eq. (1), while in the intermediate and long junctions $\Ic$ is $\propto e^{-k_B T/\dE}$. The slope of $\log{\Ic}$ vs. $T$ dependence allows us to extract the energy scale  $\dE$, which depends on the junction length but not the gate voltage $V_G$. While consistent for very long junctions $L \gg \xi$, the values of $\dE$ for intermediate devices $L \sim\xi$ are smaller than those naively estimated from the junction lengths. We attribute this suppression to the finite coherence length. Finally, we show that at the lowest temperature, $\Ic$ saturates at a level determined by the product of $\Delta_0$ or $\dE$ (depending on the regime), and the number of transversal modes across the junction width. Our observations demonstrate the universality of the critical current in several regimes relevant to most hybrid superconductor-encapsulated graphene devices.

\begin{acknowledgments} I.V.B. and  M.Y.  acknowledge the Canon foundation. C.T.K., M.T.W., A.S., and G.F. were supported by ARO Award W911NF-16-1-0122. F.A. acknowledges the ARO under Award W911NF-14-1-0349. A.W.D. was supported by the NSF graduate research fellowship DGF1106401. Low-temperature measurements performed by G.F. were supported by the Division of Materials Sciences and Engineering, Office of Basic Energy Sciences, U.S. Department of Energy, under Award DE-SC 0002765. This work was performed in part at the Duke University Shared Materials Instrumentation Facility (SMIF), a member of the North Carolina Research Triangle Nanotechnology Network (RTNN), which is supported by the National Science Foundation (Grant ECCS-1542015) as part of the National Nanotechnology Coordinated Infrastructure (NNCI). M.Y. acknowledges financial support by Grant-in-Aid for Scientific Research on Innovative Areas ``Science of Atomic Layer''. M.Y. and S.T. acknowledge support by Grant-in-Aid for Scientific Research S (No. 26220710), and Grant-in-Aid for Scientific Research A (No. 26247050). 

We would like to thank Konstantin Matveev for fruitful discussions.
\end{acknowledgments} 



\clearpage
\newpage

\begin{center}
\textbf{\large Supplementary to: Ballistic graphene Josephson junctions from the short to the long regime}
\end{center}



\beginsupplement

\section*{ Differential resistance $dV/dI (\Vg, \Ib)$ Junctions B, D, E, F, and G}
Figure S1 shows maps of the differential resistance versus bias current and gate voltage taken at base temperature for Junctions B-G. Oscillations of the critical current $I_C$ in the p-doped regime are easier to observe in Junctions A and E, which are significantly shorter than the other junctions. The dimension of the junctions are listed in Table S1.

\begin{figure}[!!!h]
\includegraphics[width=0.8 \columnwidth]{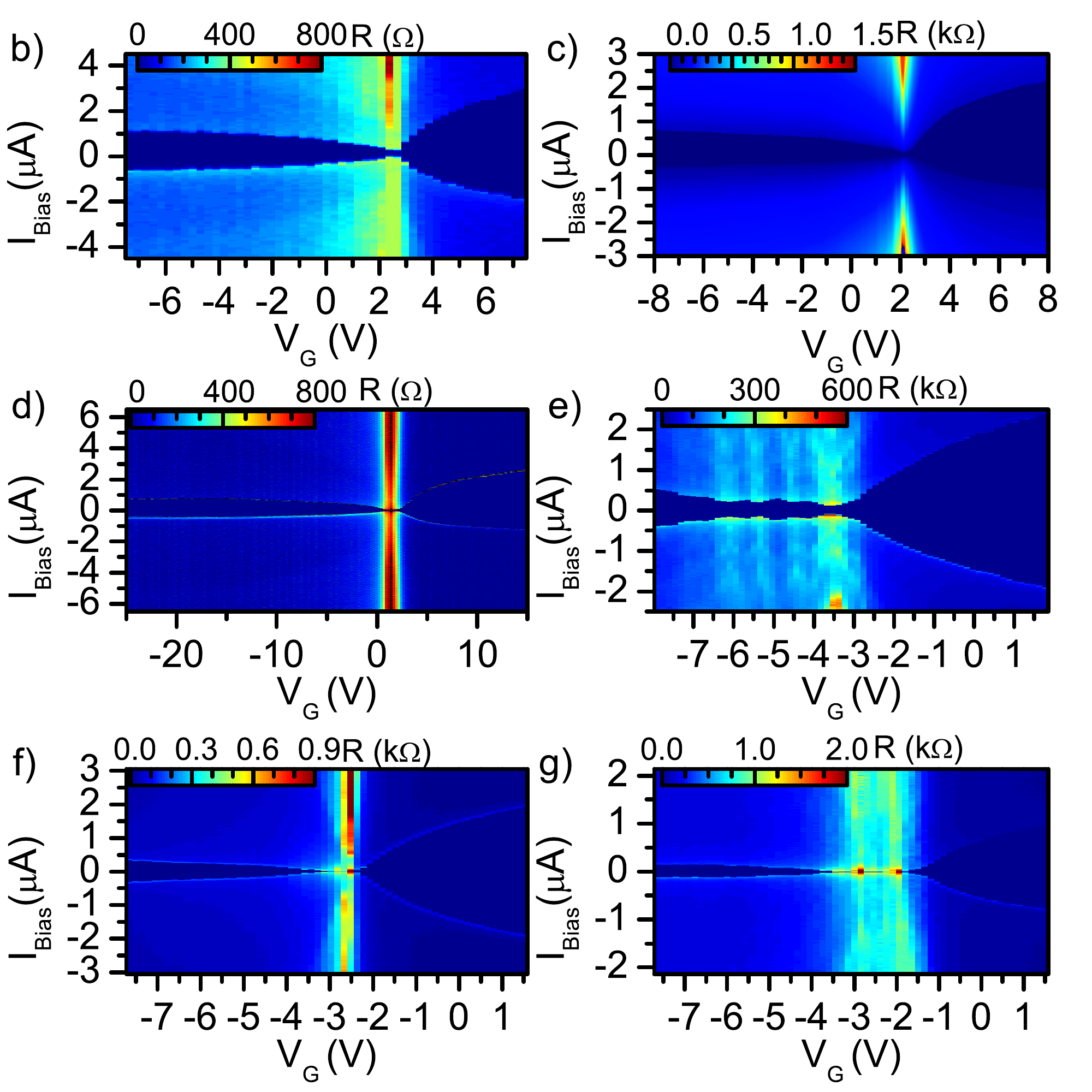}
\caption{\label{fig:overview} Maps of the differential resistance of Junctions B-G versus bias current $\Ib$ and gate voltage $\Vg$, measured at base temperatures. Around $\Ib=0$ a region of zero resistance can be observed. This superconducting region persists until $\Ib$ reaches a critical value. The current though the junction is swept from a large negative value to a large positive. Therefore, the transition at negative $\Ib$ corresponds to the retrapping current $I_R$, while the transition at positive $\Ib$ corresponds to the switching current $I_S$. 
}
\end{figure}

\begin{table*}[h!]
  \begin{center}
    \caption{List of devices}
    \label{tab:table1}
    \begin{tabular}{|c|c|c|}
      \hline
      Device & Length & Width \\ 
      \hline
      $J_{A}$  & 0.2$\,$$\mu m$ & 3.0$\,$$\mu m$  \\
      $ J_{B}$  & 0.4$\,$$\mu m$ & 3.0$\,$$\mu m$ \\
      $J_{C}$ & 1.0$\,$$\mu m$ & 3.0$\,$$\mu m$\\   
      $J_{D}$ & 2.0$\,$$\mu m$ & 5.0$\,$$\mu m$\\   
      $J_{E}$ & 0.3$\,$$\mu m$ & 2.4$\,$$\mu m$\\   
      $J_{F}$  & 0.65$\,$$\mu m$ & 4.5$\,$$\mu m$  \\
      $J_{G}$ & 0.8$\,$$\mu m$ & 2.4$\,$$\mu m$\\   
      \hline
    \end{tabular}
  \end{center}
\end{table*}

\section*{On the relationship between the measured switching current $I_S$ and the true critical current $I_C$}
Generally speaking, due to damping, environmental, and thermal effects the true critical current, $I_C$, of a small Josephson junction is inaccessible; instead a smaller switching current $I_S$ is measured. 
In particular, in evaluating the dynamics of the junctions (overdamped vs. underdamped) it is important to consider the capacitance of the large ($\sim 100\times 100\mu m$) bonding pads coupled via a global back gate, which could contribute up to $1pF$ to the junction capacitance. In our earlier work (Ref.~\onlinecite{Overheat}), we fabricated $\sim 100\Omega$  resistors in the leads in order to isolate the bonding pads, making the junctions certainly overdamped. In this work, the junctions are not purposefully isolated from the bonding pads. Therefore, for the shortest junctions and at high critical currents, we estimate that the quality factor could potentially reach up to $Q \sim 5$, the slightly underdamped regime. (The real $Q$ of the junction may either be reduced due to the lead impedance, or increased due to the effectively suppressed internal dissipation of the junction at low temperatures.) We therefore chose to verify our approximation $I_C \approx I_S$ experimentally, by analyzing the statistical distribution of $I_S$. 

The distribution is measured by repeatedly sweeping the bias current from zero past the switching, and recording $I_S$ at every sweep~\cite{Lee2011,Fulton}. Thus we obtain $P(I)$, the probability that the junction will switch from the superconducting to the normal state at bias current $I_{Bias}$. Obtaining the critical current $I_C$ via statistical methods, while more accurate, would take a prohibitively long time if performed for every junction, at every gate voltage $V_G$ and temperature $T$. Instead, we choose one of the shorter junctions (E, 300 nm long), which would more likely demonstrate premature switching due to being underdamped. We then select several representative gate voltages and verify that the measured switching current $I_S$ is very close to the true critical current $I_C$ (Figure S2a). 

\begin{figure}[h!!]
\includegraphics[width=0.5 \columnwidth]{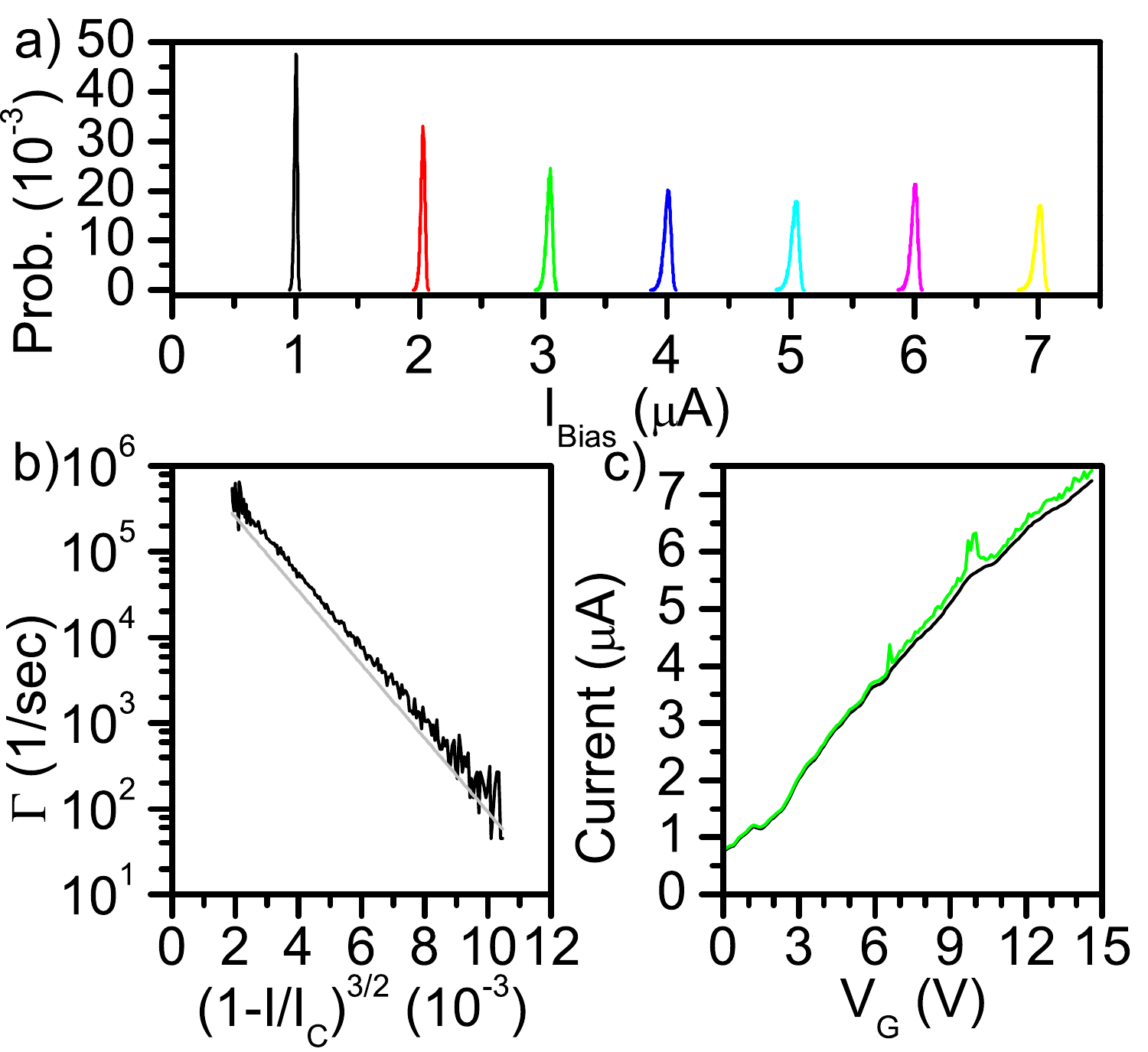}
\caption{\label{fig:overview} a) $P(I)$ histogram measured for Junction (E) at the cryostat base temperature, $T=50$ mK. Here we show data for several representative values of $V_G$. b)  The calculated escape rate $\Gamma$ plotted on a logarithmic scale versus the expression $(1-I/I_C)^{3/2}$.  c) Measured $I_S$ (black) and  fitted $I_C$ (green) versus gate voltage $V_G$. $I_S\approx I_C$ throughout the range.
}
\end{figure}

Instead of the probability distribution $P(I)$, it is more informative to look at the junction escape rate $\Gamma(I)$, which may be obtained by summing over the switching histogram~\cite{Fulton,Clarke1988}
\begin{equation}
	\Gamma(I)=\frac{dI}{dt}\frac{1}{I} \ln \left( P(I)/[1-\int \limits_{0}^{I}P(I)dI ] \right)
\end{equation} \label{equ.1}

\noindent
where $dI/dt$ is the current sweeping rate. The resulting calculation $\Gamma (I)$ allows one to estimate the true critical current $I_C$. As the bias current approaches the critical current $I_{Bias}\rightarrow I_C$, $\Gamma$ should follow the relationship: $\log(\Gamma)\propto(1-I/I_C)^{a}$, with $a=3/2$ for switching mediated by thermal activation ~\cite{Fulton, Clarke1988}. An example of this dependence is shown in Figure S2b. By fitting this dependence, we obtain a good estimate of the critical current $I_C$. We find that as $I_C$ increases, so does its difference from $I_S$. However, even for the largest presented currents ($\sim7\mu$A), the measured current is suppressed by not more than $10\%$, so we are justified in using the measured $I_S$ in place of $I_C$ in the main text. 

\section*{ $I_C$ vs. temperature $T$ behavior of Junctions E, F and G}
Figure S3 shows the critical current $I_C (T)$ on a semilogarithmic scale for devices E, F and G.  Similar to devices in the main text, $\log(\Ic)$ is clearly linear in $T$ over more than an order of magnitude, consistent with the expectation for long ballistic junctions. 

\begin{figure}[h!!]
\includegraphics[width=0.7 \columnwidth]{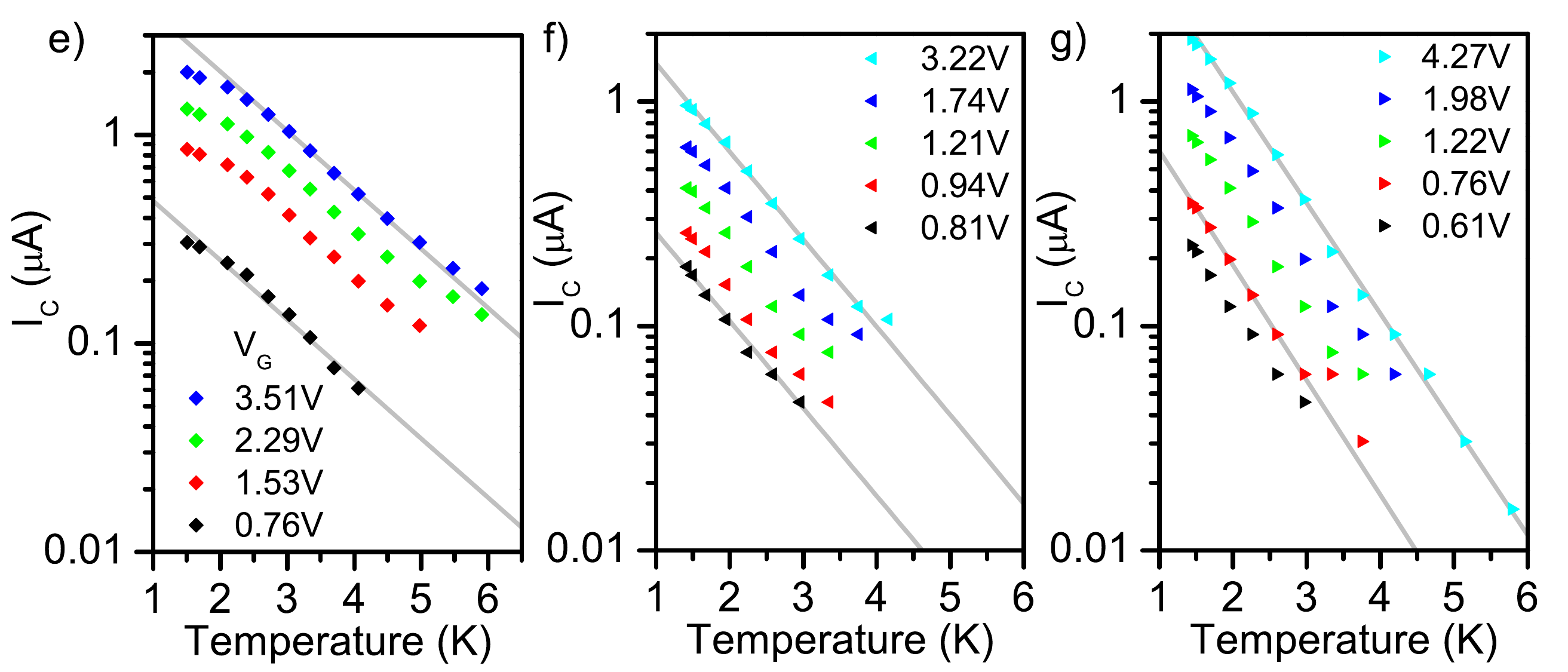}
\caption{\label{fig:overview}  Critical current $\Ic$ plotted on a semi-log scale versus temperature $T$ for e) Junction E, f) Junction F  and g) Junction G.  The slope of $\log{(\Ic)}$ vs. $T$ allows us to extract the characteristic energy $\dE$, which is almost independent of $\Vg$ (see Figure 3b of the main text.)}
\end{figure}

\begin{figure}[h!!]
\includegraphics[width=0.75 \columnwidth]{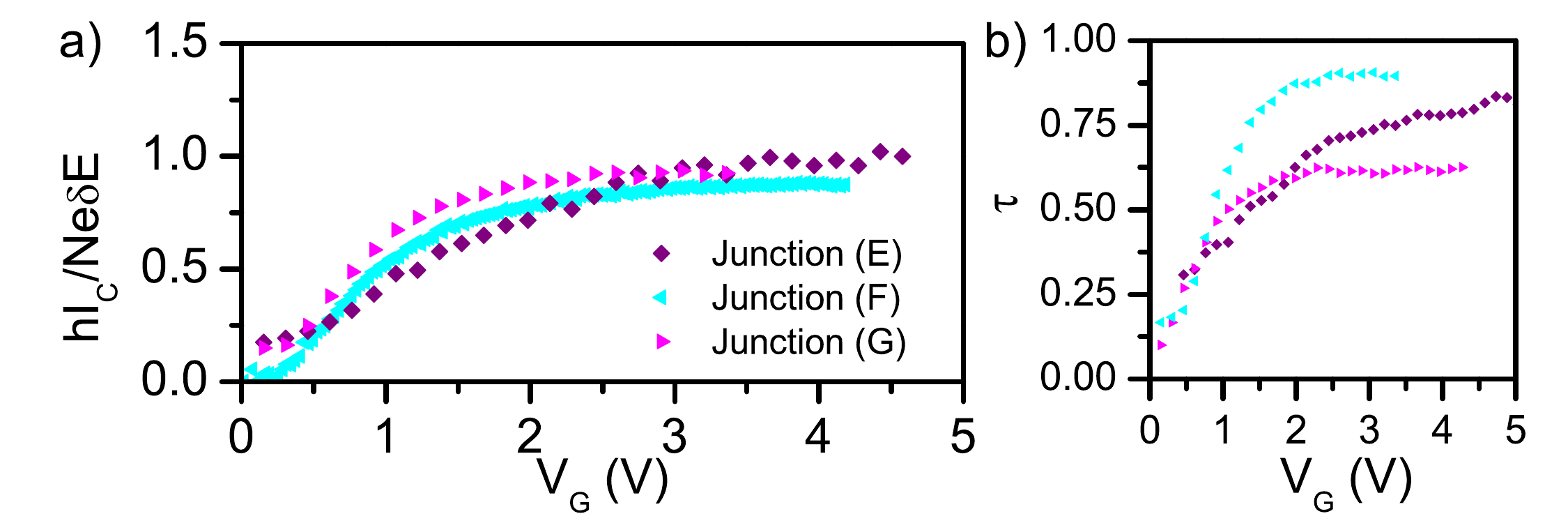}
\caption{\label{fig:overview} a)  Ratio $h\Ic/N e \dE$ for Junctions E-G, measured as a function of $\Vg$ at 60mK. At higher $V_G$ the ratio converges toward a constant value in all three junctions, consistent with devices B and C in the main text.b) Transmission coefficient $\tau$ for Junctions E-G calculated by comparing the measured conductance $G_N$ to the ballistic limit $G_0$.  }
\end{figure}

\section*{Dimensionless scaling of $I_C$ vs. $T$}
\begin{figure}[h]
\includegraphics[width=0.75 \columnwidth]{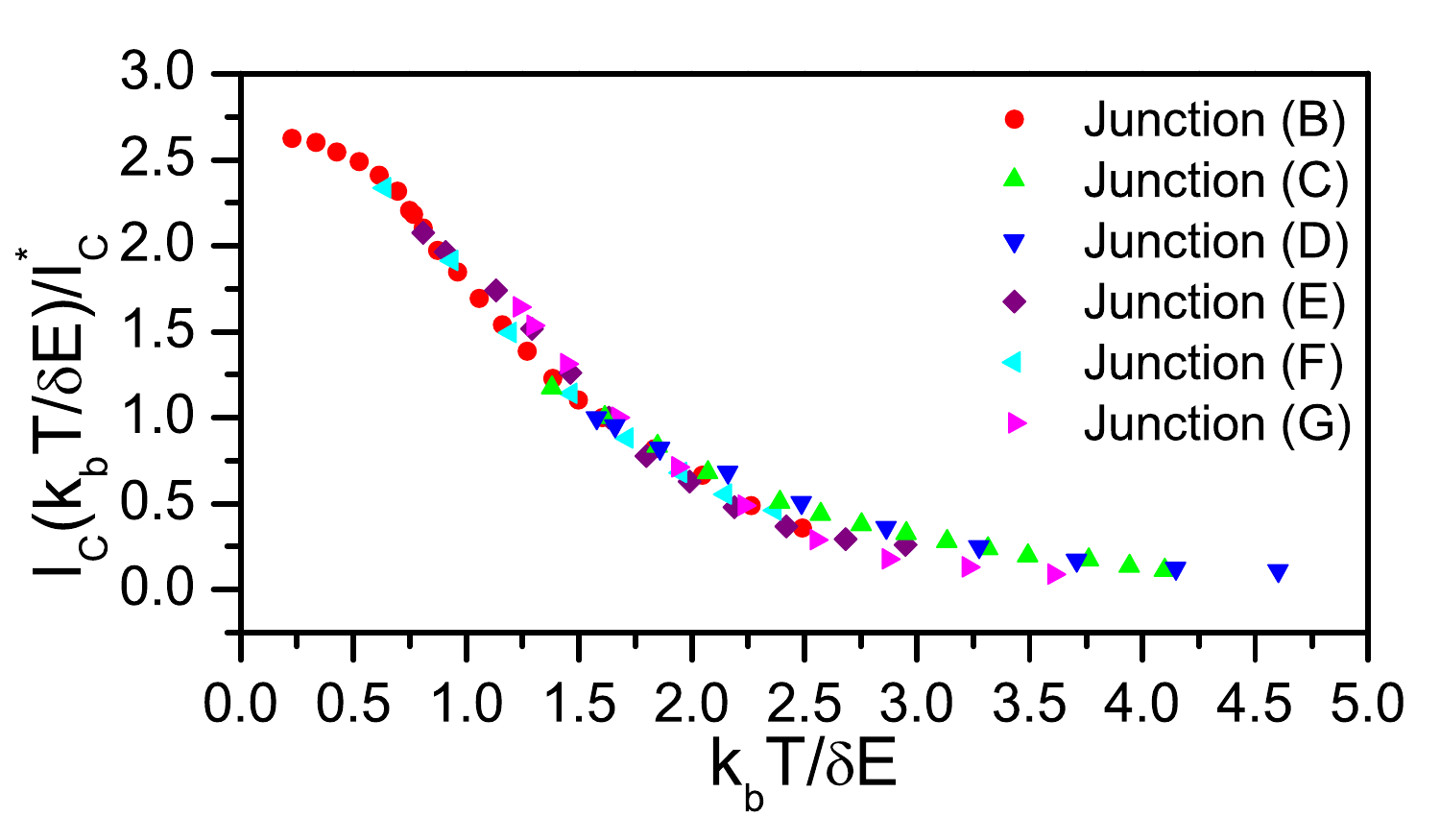}
\caption{Normalized current versus $k_B T/\delta E$ ratio demonstrating dimensionless scaling of the critical current. The critical current $I_C(k_B T/\delta E)$ is normalized by its value at $\delta E=1.6$ (denoted as $I_C^*$), the lowest point for which data are available for all presented junctions.  (Different colors represent data for different devices.) Accounting for length ($\dE$) and carrier concentration ($I_C^*$) makes the data fall on the same universal curve.}
\end{figure}

Knowing the energy scale $\dE$, we can examine the universality of the critical current $I_C$ in these ballistic devices. Instead of the absolute temperature scale, we use the device-independent ratio $k_B T/\dE$, which accounts for the difference in junction lengths. Next, we normalize the magnitude of the critical current, thus accounting for the device width, as well as the carrier concentration. The normalization point is chosen at $k_B T/\dE=1.6$, the lowest point for which data are available for all three junctions. The normalized current plotted versus $k_B T/\dE$ falls on the same curve for the four junctions and different gate voltages (Figure S5), which strongly indicates that the critical current is a universal function of the dimensionless ratio $k_B T/\dE$.

\section*{ The effect of $\tau$ on the $I_C$ vs. $\delta E$ relationship at low $T$}

We observed that at high density and $T$ approaching zero, each of the $N$ transverse modes contributes a supercurrent on the order of $e\delta E/h$.  This is illustrated in Figure 4 of the main paper, by showing that the ratio $h I_C /Ne\delta E = eI_C R_{SH}/\delta E$ tends to a constant at high $V_G$. However this ratio is suppressed close to charge neutrality, which we attributed to the lower contact transmission close to the Dirac point, as illustrated in Figure 3a inset. 

In Figure S7 we plot the ratio of $eI_CR_N/\tau\delta E$ to account for the fact that the normal resistance $R_N$ slightly differs from $R_{SH}$ and the supercurrent carried by each mode should be reduced as $\tau e\delta E/h$. Here we observe that the ratio saturates and becomes constant at gate voltages much closer to the Dirac point, which suggests that the reduction in $I_CR_N$ is indeed caused by the imperfect transmission at charge neutrality.

\begin{figure}[h!!]
\includegraphics[width=0.42\columnwidth]{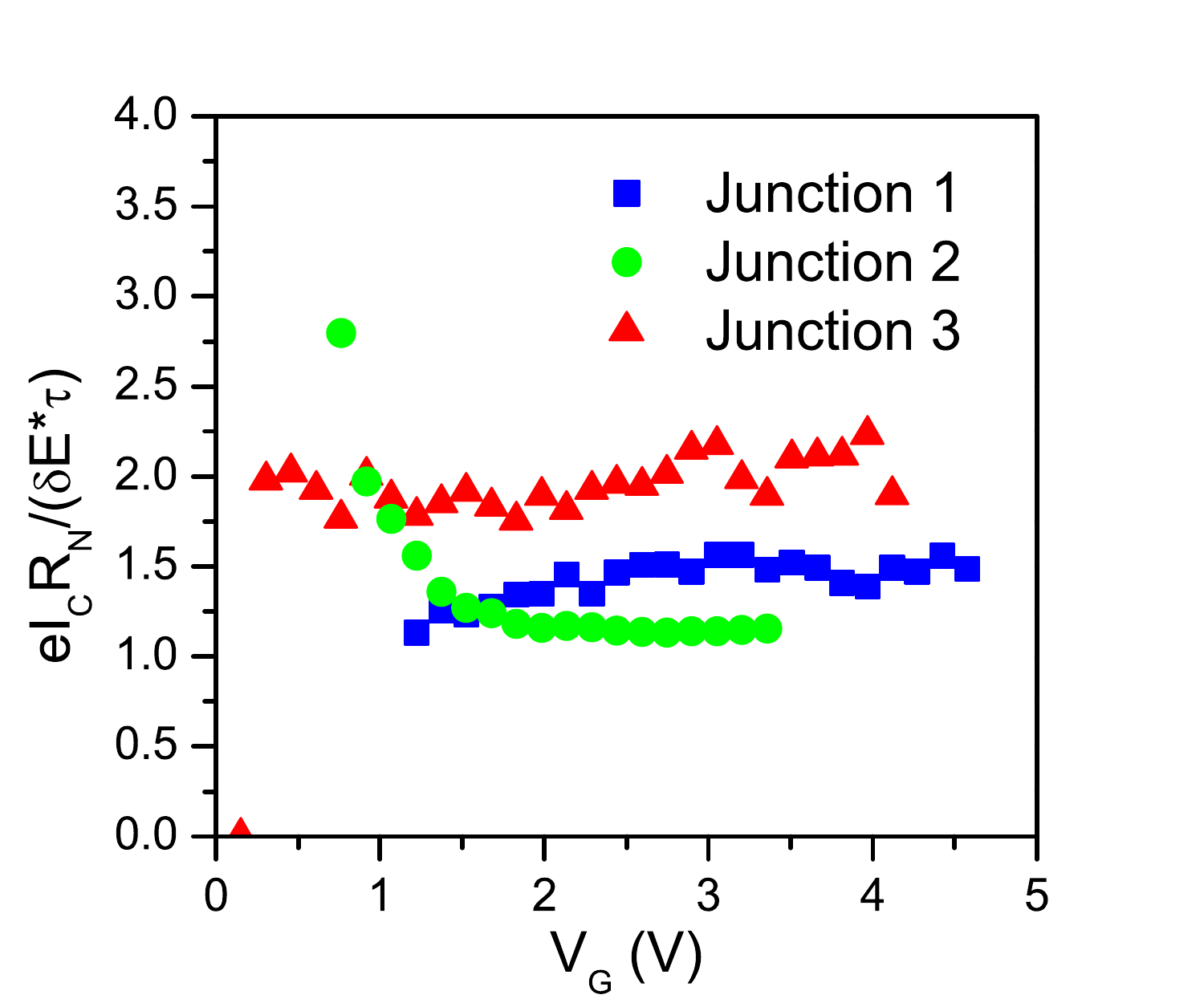}
\caption{\label{fig:overview}The ratio $eI_CR_N/\tau\dE$ for Junction B (blue $\Box$), E (red $\bigtriangleup$), and F (green $\bigcirc$), which is nearly independent of gate voltage for each junction.
}
\end{figure}


\begin{thebibliography}{99}

\bibitem{Vandersypen2015}
V. E. Calado, S. Goswami, G. Nanda, M. Diez, A. R. Akhmerov, K. Watanabe, T. Taniguchi, T. M. Klapwijk, and L. M. K. Vandersypen, Nature Nano. \textbf{10}, 761-764 (2015).

\bibitem{Geim2015}
M. Ben Shalom, M. J. Zhu, V. I. Fal'ko, A. Mishchenko, A. V. Kretinin, K. S. Novoselov, C. R. Woods, K. Watanabe, T. Taniguchi, A. K. Geim, and J. R. Prance, Nature Phys. \textbf{12}, 318-322 (2016).

\bibitem{Yacoby}M. T. Allen, O. Shtanko, I. C. Fulga, J. I. J. Wang, D. Nurgaliev, K. Watanabe, T. Taniguchi, A. R. Akhmerov, P. Jarillo-Herrero, L. S. Levitov, and A. Yacoby, Nature Phys. \textbf{12}, 128-133 (2016).

\bibitem{amet2016}F. Amet, C. T. Ke, I. V. Borzenets, Y. Wang, K. Watanabe, T. Taniguchi, R. S. Deacon, M. Yamamoto, Y. Bomze, S. Tarucha, and Finkelstein,  Science \textbf{352} (6288), p.966 (2016).


\bibitem{Kulik} I.O. Kulik, Sov. Phys. JETP \textbf{30}, 944 (1970).

\bibitem{Bardeen}J. Bardeen and J. L. Johnson, Phys. Rev. B \textbf{5}, 72 (1972).

\bibitem{book}A. V. Svidzinskii, {\it Spacially-Inhomogeneous Problems of Theory of Superconductivity} (Nauka, Moscow, 1982).

\bibitem{Tinkham} M. Tinkham, {\it Introduction To Superconductivity} (McGraw-Hill, New York, 1996).

\bibitem{Golubov} A. A. Golubov, M. Yu. Kupriyanov, and E. Iliichev, Rev. Mod. Phys. \textbf{76}, 411 (2004).



\bibitem{Svidzinski1}A. V. Svidzinsky, T. N. Antsygina, and E. N. Bratus, Sov. Phys. JETP \textbf{3}, 860 (1972).

\bibitem{Svidzinski2}A. V. Svidzinsky, T. N. Antsygina, and E. N. Bratus,  J. Low Temp. Phys. \textbf{10}, 131-136 (1973).


\bibitem{Beenakker92} C. W. J. Beenakker, {\it Transport Phenomena in Mesoscopic Systems} pp.235-253 (Springer, Berlin, 1992).

\bibitem{Beenakker91} C. W. J. Beenakker and H. van Houten, Phys. Rev. Lett. \textbf{66}, 3056 (1991).


\bibitem{Lee}G.-H. Lee, S. Kim, S.-H. Jhi, and H.-J. Lee, Nature Commun. \textbf{6}, 6181 (2015).





\bibitem{Wang} L. Wang, I. Meric, P. Y. Huang, Q. Gao, Y. Gao, H. Tran, T. Taniguchi, K. Watanabe, L. M. Campos, D. A. Muller, J. Guo, P. Kim, J. Hone, K. L. Shepard, and C. R. Dean, Science \textbf{342}, 614-617 (2014).

\bibitem{suppl}See Supplementary Information

\bibitem{heersche_2007}H. B. Heersche, P. Jarillo-Herrero, J. B. Oostinga, L. M. K. Vandersypen, and A. F. Morpurgo, Nature {\bf 446}, 56 (2007).

\bibitem{du_2008}X. Du, I. Skachko, and E. Y. Andrei,  Phys. Rev. B {\bf 77}, 184507 (2007).

\bibitem{gueron_2009}C. Ojeda-Aristizabal, M. Ferrier, S. Gu\'{e}ron, and H. Bouchiat, Phys. Rev. B {\bf 79}, 165436 (2009).

\bibitem{23}  I. V. Borzenets, U. C. Coskun, S. J. Jones, and  G. Finkelstein, Phys. Rev. Lett. {\bf 107}, 137005 (2011).

\bibitem{24} I. V. Borzenets, U. C. Coskun, S. J. Jones, and G. Finkelstein, IEEE Trans. Appl. Supercond. {\bf 22}, 1800104 (2012).

\bibitem{Overheat} I.V. Borzenets, U.C. Coskun, H.T. Mebrahtu, Yu.V. Bomze, A.I. Smirnov, and G. Finkelstein, Phys. Rev. Lett. \textbf{111}, 027001 (2013).

\bibitem{Pekola} H. Courtois, M. Meschke, J. T. Peltonen, and J. P. Pekola, Phys. Rev. Lett. \textbf{101}, 067002 (2008).

\bibitem{Ulas}U.C. Coskun, M. Brenner, T. Hymel, V. Vakaryuk, A. Levchenko, and A. Bezryadin, Phys. Rev. Lett. \textbf{108}, 097003 (2012).

\bibitem{HJ_Lee}G.-H. Lee, D. Jeong, J.H. Choi, Y.-J. Doh, and H.-J. Lee, Phys. Rev. Lett. \textbf{107}, 146605 (2011).

\bibitem{Ting}C. T. Ke, I. V. Borzenets, A. W. Draelos, F. Amet, Yu. Bomze, G. Jones, M. Craciun, S. Russo, M. Yamamoto, S. Tarucha and G. Finkelstein, Nano Lett. \textbf{16} (8), 4788 (2016).

\bibitem{Young}A. F. Young, and P.  Kim,  Nature Phys. \textbf{5}, 222-226 (2009).

\bibitem{Rickhaus}P. Rickhaus, R. Maurand, M.-H. Liu, M. Weiss, K. Richter, and C. Schönenberger, Nature Commun. \textbf{4}, 2342 (2013).

\bibitem{Hagymasi}I. Hagymasi, A. Kormanyos, and J. Cserti, Phys. Rev. B \textbf{82}, 134516 (2010).

\bibitem{GapT}R. Dougherty, J. D. Kimel, {\it Superconductivity Revisited}, (CRC Press, London 2013).


\bibitem{Bagwell} P. F. Bagwell,  Phys. Rev. B \textbf{46}, 12573-12586 (1992).


\bibitem{Ishii}C. Ishii, Prog. Theor. Phys. \textbf{44}, 1525 (1970).





\bibitem{Titov}M. Titov and C. W. J. Beenakker, Phys. Rev. B \textbf{74}, 041401(R) (2006).







%

%


\end{thebibliography}

\begin{thebibliography}{99}

\bibitem{Overheat} I.V. Borzenets, U.C. Coskun, H.T. Mebrahtu, Yu.V. Bomze, A.I. Smirnov, and G. Finkelstein, Phys. Rev. Lett. \textbf{111}, 027001 (2013).


\bibitem{Lee2011}G.-H. Lee, D. Jeong, J.-H. Choi, Y.-J. Doh, and H.-J. Lee, Phys. Rev. Lett. {\bf 107}, 146605 (2011).

\bibitem{Fulton}T. A. Fulton and L. N. Dunkleberger, Phys. Rev. B {\bf 9}, 4760 (1974).

\bibitem{Clarke1988}J. Clarke, E. A. Cleland, H. M. Devoret, D. Esteve, and M. J. Martinism, Science {\bf239}, (1988).
\bibitem{Golubov}A. A. Golubov, M. Yu. Kupriyanov, and E. Iliichev, Rev. Mod. Phys \textbf{76}, 411 (2004).

\bibitem{Kuprianov}M. Yu. Kuprianov and V. F. Lukichev, Zh. Eksp. Teor. Fiz. \textbf{94}, 139-149 (1988).

\bibitem{Bagwell}P. F. Bagwell, Phys. Rev. B \textbf{46}, 12573-12586 (1992).

\bibitem{Shiba}Y. Luh, Acta Phys. Sin. \textbf{21}, 75 (1965); Shiba, H. Classical spins in superconductors. {\it Prog. Theor. Phys.} \textbf{40}, 435 (1968); A. I. Rusinov, A. I. On the theory of gapless superconductivity in alloys containing paramagnetic impurities. {\it Sov. Phys. JETP} \textbf{29}, 1101 (1969). 

\bibitem{Eilenberger} Eilenberger, G. Transformation of Gorkov’s Equation for type II superconductors into transport-like equations. {\it Z. Phys.} \textbf{214}, 195-213 (1968).

\end{thebibliography}
\end{document}